# Random Phase Product Sate for Canonical Ensemble

Toshiaki IITAKA (飯高敏晃)
*RIKEN Center for Computational Science (R-CCS),
2-1 Hirosawa, Wako, Saitama, 351-0198, Japan.
e-mail address: tiitaka@riken.jp*
URL: http://www.iitaka.org/rpps/

Method of *random phase product state* (RPPS) is proposed to calculate canonical ensemble average of quantum systems described with matrix product states and also with tensor network states in general. The RPPS method is an extension of the method of *random phase state* for full Hilbert space representation. The validity of the method is confirmed by comparing the average energy of N-site antiferromagnetic spin-1/2 Heisenberg chain model with open boundary conditions with the result of direct method (for up to N=14) and *minimally entangled typical thermal state* (METTS) method (for N=100). Numerical advantages of the RPPS such as parallelization, combined calculation of thermal averages at different temperatures, parameters for controlling error are discussed. View point of *self-averaging* for the super-convergence of random state method is emphasized in addition to that of *typicality*.

According to the textbook of statistical mechanics[1], physical quantities of the system in thermal equilibrium at inverse temperature $\beta = 1/k_B T$ are calculated based on the canonical ensemble, in which each microscopic quantum state $\phi_n$ with energy $E_n$ is realized according to the Gibbs probability $P_n \propto e^{-\beta E_n}$. Calculating physical quantities with canonical ensemble of quantum many body system is one of the most important problems in computational physics, which requires, however, formidable computational resources if direct diagonalization method is applied to large systems.

There are two sources of the difficulty. First, the dimension of the state vector increases exponentially as the system size increases. For example, state vector of the spin one-half Heisenberg chain has dimension of $M = 2^N$ where N is the number of spins [2]. This difficulty may be mitigated by using tensor network representation [3, 4] of quantum states, which efficiently represents physically meaningful states near the ground states using much less parameters than conventional full Hilbert space representation.

Second, the number of excited states to be included in the calculation increases exponentially as temperature increases. [1]. In contrast, at very low temperature, only few states close to the ground states contribute to the thermal average and these low lying states are efficiently calculated with Lanczos methods [5, 6]. To cope with the second difficulty, method of random state has been successfully applied to many problems (see [6] and references therein) and also discussed in terms of typicality[7-12]. Among them, method of *random phase state* [13-18] has superior property of smaller statistical fluctuation. The main idea of random (phase) state method is sampling the full Hilbert space with much smaller number of random states than the full basis set. So far, the method of random state has been used with full Hilbert space representation and only few authors used it with tensor network representation[7-10].

In the following, the method of random state is reviewed and extended to tensor network representation using, as an example, matrix product state (MPS) [3, 4] of N-site anti-ferromagnetic spin-1/2 Heisenberg chain model with open boundary condition. However, extension to tensor network in general is straight forward.

The Hamiltonian of antiferromagnetic N-site spin-1/2 Heisenberg chain with open boundary condition is defined as

$$H = \sum_{i=1}^{N-1} \vec{S}_i \cdot \vec{S}_{i+1} \qquad (1)$$

where $\vec{S}_i$ is spin-1/2 operator at site *i*.

A quantum state of this system is expressed in full Hilbert space representation as

$$|\phi\rangle = \sum_{\sigma_1 \cdots \sigma_N} c^{\sigma_1 \cdots \sigma_N} |\sigma_1 \cdots \sigma_N\rangle = \sum_{\boldsymbol{\sigma}} c^{\boldsymbol{\sigma}} |\boldsymbol{\sigma}\rangle \qquad (2)$$

where the tensor of rank N, $c^{\boldsymbol{\sigma}} = c^{\sigma_1 \cdots \sigma_N}$, is the components of $|\phi\rangle$ projected onto the complete orthonormal basis set

$$|\sigma_1 \cdots \sigma_N\rangle = |\sigma_1\rangle \otimes |\sigma_2\rangle \otimes \cdots |\sigma_N\rangle \qquad (3)$$

constructed as the direct product of eigenstates of local spin operator

$$S_{zi}|\sigma_i\rangle = \frac{\hbar}{2}\sigma_i|\sigma_i\rangle \qquad (4)$$

with eigenvalues $\sigma_i = \pm 1$.

In order to sample the full Hilbert space, *random state* is defined [16] as

$$|\Phi\rangle = \frac{1}{\sqrt{M}} \sum_{\boldsymbol{\sigma}} \xi^{\boldsymbol{\sigma}} |\boldsymbol{\sigma}\rangle \quad (5).$$

where $\xi^\sigma$ is a set of *independent and identically distributed* (*i.i.d.*) complex random numbers that have the following statistical relations

$$\langle\langle \xi^\sigma \rangle\rangle = 0 \quad (6)$$

$$\langle\langle \xi^{\sigma'*} \xi^\sigma \rangle\rangle = \delta_{\sigma'\sigma} \quad (7)$$

$$\langle\langle \xi^{\sigma'} \xi^\sigma \rangle\rangle = 0 \quad (8)$$

where the double bracket and the asterisk indicate statistical average and complex conjugate, respectively.

As a special case of such random numbers, random phase number is defined as

$$\xi^\sigma = \exp[i\theta^\sigma] \quad (9)$$

where $\theta^\sigma$ are M independent uniform real random numbers in the range $[0, 2\pi)$ where M=$2^N$ is the dimension of the full Hilbert space. Random state composed of random phase numbers is called *random phase state* [14, 16].

Then, normalization and completeness of random state are expressed as

$$\langle\langle \langle \Phi | \Phi \rangle \rangle\rangle = 1 \quad (10)$$

or

$$\langle \Phi | \Phi \rangle = 1 \quad (11)$$

in the special case of random phase state, and

$$\langle\langle |\Phi\rangle\langle\Phi| \rangle\rangle = \frac{I}{M} \quad (12)$$

respectively. Matrix element of operator $X$ gives the trace of $X$,

$$\langle\langle \langle \Phi | X | \Phi \rangle \rangle\rangle = \frac{1}{M}\sum_{\sigma'\sigma}\langle\langle \xi^{\sigma'*}\xi^\sigma \rangle\rangle X_{\sigma'\sigma}$$
$$= \frac{1}{M}\sum_\sigma X_{\sigma\sigma} = \frac{1}{M} tr[X]$$
(13)

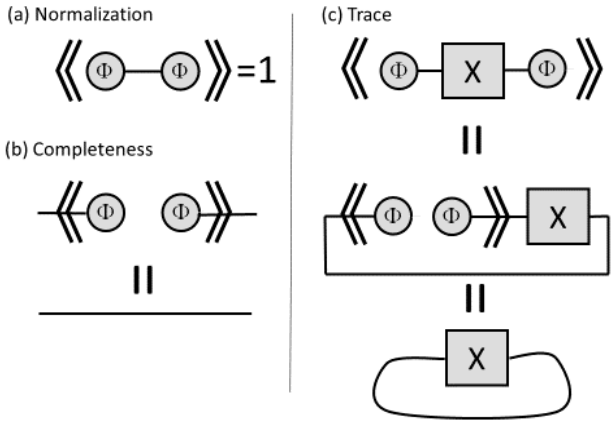

Fig. 1. Graphic representation of random state method: (a) normalization (10); (b) completeness (12); (c) trace

(13). The double brackets indicate statistical average. Normalization factor of 1/M is omitted for clarity.

Thermal state at inverse temperature $\beta$ is defined as

$$|\Phi(\beta)\rangle = e^{-\beta H/2}|\Phi\rangle \quad (14)$$

Thermal average of physical quantity *A* with canonical ensemble is calculated as

$$\langle A \rangle = \frac{Tr[e^{-\beta H}A]}{Tr[e^{-\beta H}]} = \frac{\langle\langle \langle \Phi(\beta)|A|\Phi(\beta)\rangle \rangle\rangle}{\langle\langle \langle \Phi(\beta)|\Phi(\beta)\rangle \rangle\rangle}$$
(15)

The essence of the random state method is that the double summation in expectation value (13) is reduced to single summation in trace by using statistical relation (7) as illustrated in Fig.1.

When statistical average $\langle\langle \langle \Phi|X|\Phi\rangle \rangle\rangle$ is calculated as *sample average* with $N_{sample}$ realizations of random states, the fluctuation of the average decreases as

$$\frac{|\delta X|}{\sqrt{N_{sample}}}$$

according to the law of large numbers. The statistical fluctuation of $\langle \Phi|X|\Phi\rangle$ for each realization of random state is given by

$$\delta X = \langle \Phi|X|\Phi\rangle - \langle\langle \langle \Phi|X|\Phi\rangle \rangle\rangle$$
$$= \frac{1}{M}\sum_{\sigma_1\sigma_2}(\xi^{\sigma_1*}\xi^{\sigma_2} - \delta_{\sigma_1\sigma_2})X_{\sigma_1\sigma_2}$$
(16)

Therefore the magnitude of the fluctuation is expected as

$$|\delta X|^2 = \frac{1}{M^2}\sum_{\sigma_1\sigma_2\sigma_3\sigma_4}\langle\langle (\xi_{\sigma_1}\xi^*_{\sigma_2} - \delta_{\sigma_1\sigma_2})X^*_{\sigma_1\sigma_2} \times$$
$$(\xi^*_{\sigma_3}\xi_{\sigma_4} - \delta_{\sigma_3\sigma_4})X_{\sigma_3\sigma_4} \rangle\rangle$$
$$= \frac{1}{M^2}\sum_{\sigma_1\neq\sigma_2}|X_{\sigma_1\sigma_2}|^2 + \frac{1}{M^2}\left(\langle\langle |\xi_\sigma|^4 \rangle\rangle - 1\right)\sum_\sigma |X_{\sigma\sigma}|^2$$
(17)

According to the inequality

$$\langle\langle |\xi_\sigma|^4 \rangle\rangle \geq \langle\langle |\xi_\sigma|^2 \rangle\rangle^2 = 1$$

the second term of (17) becomes zero if and only if $|\xi_\sigma| = 1$ for all $\sigma$. Since this condition is satisfied by random phase state, it gives the smallest fluctuation among the random states in the given basis set. In many cases such as combinations of local operators, $X_{\sigma_1\sigma_2}$ is *sparse* (i.e. most of the matrix elements are zero) and the right hand side of (17) decreases rapidly as $\frac{1}{M} \propto 2^{-N}$ as system size N increases. This is *self-averaging*[16, 19].

Now let us extend the method of random state to tensor network representation. Here, Matrix Product State (MPS), a special form of tensor network decomposition of $c^\sigma$, useful for one dimensional system, is used as an example. MPS is defined as

$$\begin{aligned}|\phi\rangle &= \sum_\sigma c^\sigma |\boldsymbol{\sigma}\rangle \\ &= \sum_\sigma \left[A^{[1]\sigma_1}\right]\left[A^{[2]\sigma_2}\right]\cdots\left[A^{[N-1]\sigma_{N-1}}\right]\left[A^{[N]\sigma_N}\right]|\boldsymbol{\sigma}\rangle \\ &= \sum_\sigma \sum_{\substack{\mu_1\cdots\mu_{N-1} \\ \mu_0=\mu_N=1}} A^{[1]\sigma_1}_{\mu_0\mu_1} A^{[2]\sigma_2}_{\mu_1\mu_2} \cdots A^{[N-1]\sigma_{N-1}}_{\mu_{N-2}\mu_{N-1}} A^{[N]\sigma_N}_{\mu_{N-1}\mu_N} |\boldsymbol{\sigma}\rangle\end{aligned}$$
(18)

where $c^\sigma$ is decomposed into the product of N matrices $\left[A^{[i]\sigma}\right]$ of dimension $\chi$ or *bond dimension[3, 4]*. The bond index $\mu_i$ runs from 1 to $\chi$ with constraint $\mu_0 = \mu_N = 1$ due to the open boundary conditions. The $M = 2^N$ coefficients of $c^\sigma$ in full Hilbert space representation is reduced to $\sim 2N\chi^2$ coefficients in MPS. If $\chi$ were exponentially large (i.e. $2N\chi^2 \approx 2^N$), MPS would have sufficient flexibility to express the exact state [4].

Random matrix product state is defined as

$$|\Psi\rangle = \frac{1}{\sqrt{M}} \sum_\sigma \eta^\sigma |\boldsymbol{\sigma}\rangle \quad (19).$$

where coefficient $\eta^\sigma$ is defined by

$$\eta^\sigma = \chi^{-(N-1)/2} \sum_{\substack{\mu_1\cdots\mu_{N-1} \\ \mu_0=\mu_N=1}} \xi^{[1]\sigma_1}_{\mu_0\mu_1} \xi^{[2]\sigma_2}_{\mu_1\mu_2} \cdots \xi^{[N-1]\sigma_{N-1}}_{\mu_{N-2}\mu_{N-1}} \xi^{[N]\sigma_N}_{\mu_{N-1}\mu_N}$$
(20)

with *i.i.d.* complex random numbers, $\xi^{[i]\sigma}_{\mu\nu}$, that satisfy the statistical relations,

$$\left\langle\!\left\langle \xi^{[i]\sigma}_{\mu\nu} \right\rangle\!\right\rangle = 0 \quad (21)$$

$$\left\langle\!\left\langle \xi^{[i]\sigma*}_{\mu'\nu'} \xi^{[j]\sigma}_{\mu\nu} \right\rangle\!\right\rangle = \delta_{ij}\delta_{\sigma'\sigma}\delta_{\mu'\mu}\delta_{\nu'\nu} \quad (22)$$

$$\left\langle\!\left\langle \xi^{[i]\sigma'}_{\mu'\nu'} \xi^{[j]\sigma}_{\mu\nu} \right\rangle\!\right\rangle = 0 \quad (23).$$

Using (20)-(23), it is shown that $\eta^\sigma$ satisfies statistical relations corresponding to (6)-(8),

$$\left\langle\!\left\langle \eta^\sigma \right\rangle\!\right\rangle = 0 \quad (24)$$

$$\left\langle\!\left\langle \eta^{\sigma'*} \eta^\sigma \right\rangle\!\right\rangle = \delta_{\sigma'\sigma} \quad (25)$$

$$\left\langle\!\left\langle \eta^{\sigma'} \eta^\sigma \right\rangle\!\right\rangle = 0 \quad (26)$$

although $M = 2^N$ random numbers of $\eta^\sigma$ are *not* independent of each other because they are composed of only $2N\chi^2 \ll 2^N$ independent random numbers, $\xi^{[i]\sigma}_{\mu\nu}$.

Therefore random matrix product state also satisfies all equations obtained for the full Hilbert space representation. Especially, thermal MPS state is defined as

$$|\Psi(\beta)\rangle = e^{-\beta H/2}|\Psi\rangle \quad (27)$$

and the thermal average of physical quantity $A$ is expressed as

$$\langle A \rangle = \frac{Tr[e^{-\beta H}A]}{Tr[e^{-\beta H}]} = \frac{\left\langle\!\left\langle \langle\Psi(\beta)|A|\Psi(\beta)\rangle \right\rangle\!\right\rangle}{\left\langle\!\left\langle \langle\Psi(\beta)|\Psi(\beta)\rangle \right\rangle\!\right\rangle}$$
(28)

where $A$ and $H$ are matrix product operators (MPO) [4].

To extend the above results for MPS to tensor network in general, Random Phase Product State (RPPS) is defined as the random matrix product state where $\chi = 1$ in (20), so that it becomes a product state which is independent of the topology of tensor network. Topology of tensor network matters when the operators such as $A$ and $e^{-\tau_0 H/2}$ are applied to the state. In addition, random phase number (9) is used for the coefficients $\xi^{[i]\sigma_i}_{11} = \exp\left[i\theta^{[i]\sigma_i}\right]$ in (20), so that statistical fluctuation (17) becomes minimum. Therefore RPPS is defined as

$$|\Psi\rangle = \frac{1}{\sqrt{M}} \sum_\sigma \exp\left(i\theta^{[1]\sigma_1}\right)\exp\left(i\theta^{[2]\sigma_2}\right)\cdots\exp\left(i\theta^{[N]\sigma_N}\right)|\boldsymbol{\sigma}\rangle$$
(29)

RPPS can be regarded as the random matrix product state (RMPS) [7] with bond dimension one and is analogous to classical product state (CPS) in minimally entangled typical thermal state method (METTS) [20, 21].

In numerical calculation, the canonical ensemble average (28) is calculated as shown in Fig. 3 (a). Manipulation of MPS is taken care of by numerical library ITensor [22]. First, an initial RPPS is generated. Second, the thermal state $\left|\Psi^{(i)}(\beta)\right\rangle = \left[e^{-\tau_0 H/2}\right]^L \left|\Psi^{(i)}\right\rangle$ is calculated by repeatedly applying Boltzmann MPO [23], $e^{-\tau_0 H/2}$, with small imaginary time step $\tau_0 = \beta/L$, where $L$ is a large integer. Entanglement may be increased by the application of operators and the resulting MPS may require larger bond dimensions to satisfy the required cutoff error condition. Such automatic control of bond dimension is taken care of by iTensor. The *shifted and scaled* Hamiltonian

$$\bar{H} = a(H - c) \quad (30)$$

is used in program for numerical convenience. The constants $a>0$ and $c$ are determined so that all eigenvalues of $\bar{H}$ lie in the range $[0, -1]$. Third, expectation value of the observable $A^{(i)} = \left\langle\Psi^{(i)}(\beta)\right|A\left|\Psi^{(i)}(\beta)\right\rangle$ and partition function

$Z^{(i)} = \langle \Psi^{(i)}(\beta) | \Psi^{(i)}(\beta) \rangle$ are calculated. After averaging with $N_{sample}$ initial random states, the canonical average of the observable at inverse temperature $\beta$ is obtained as

$$\langle A \rangle_\beta = \frac{\sum_{i=1}^{N_{sample}} A^{(i)}}{\sum_{i=1}^{N_{sample}} Z^{(i)}} = \frac{\langle\langle A^{(i)} \rangle\rangle}{\langle\langle Z^{(i)} \rangle\rangle}.$$

The accuracy of RPPS method is controlled by the number of initial RPPS, $N_{sample}$, scaled imaginary time step $\tau_0$, cutoff parameter of SVD, $\varepsilon_{cutoff}$ [19]. The accuracy of RPPS method for the Heisenberg chain up to $N_{spin} = 14$ with $N_{sample} = 10000$, $\tau_0 = 0.02$, $\varepsilon_{cutoff} = 10^{-13}$ has been confirmed within statistical error of $10^{-4}$ in comparison with the results of the direct method[24]. The accuracy of $N_{spin} = 100$ calculation with control parameters $N_{sample} = 100$, $\tau_0 = 0.02$, $\varepsilon_{cutoff} = 10^{-10}$ was confirmed within statistical error of $10^{-4}$ in comparison with the result of METTS method using the same control parameters. In Fig. 2, the average energy and the maximum bond dimension are shown as a function of inverse temperature. It is noted that relatively small bond dimension of $\chi \sim 10$ is sufficient for converged results.

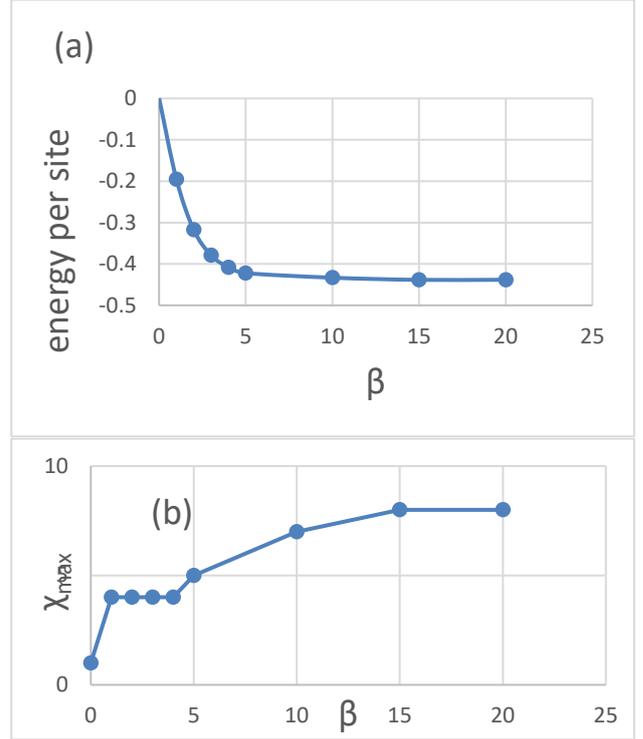

Fig. 2. Temperature dependence of (a) average energy, and (b) maximum bond dimension for 100 site spin-1/2 antiferromagnetic Heisenberg chain with open boundary condition and control parameters $N_{sample} = 100$, $\tau_0 = 0.02$, $\varepsilon_{cutoff} = 10^{-10}$.

Many other numerical methods are known for thermal calculation of tensor network state. Among them, Matrix Product Purification method [19, 25, 26] utilizes augmented sites (or ancilla) and poorly scale to large systems at low temperature. Random Matrix Product State (RMPS) studied by Garnerone [7] reduces to RPPS when $\chi = 1$. Minimally Entangled Typical Thermal State (METTS) [20, 21] scales to large systems at low temperature. The similarity of algorithm of RPPS and METTS are demonstrated in Fig. 3. In this sense, RPPS is algorithm of new generation that has inherited gene from both RMPS and METTS. RPPS has better parallelization efficiency than METTS because it has no sequential loop. In addition, RPPS can calculate thermal averages at intermediate inverse temperatures $\beta_l = l\tau_0$, where $l = 1, 2, 3, \cdots, L-1$, as the byproducts of the calculation at $\beta = L\tau_0$. Unlike METTS, RPPS does not have problems related to the ergodicity of sampling because it calculates the trace over the entire Hilbert space as shown in (13). RPPS starts with an initial product state and the bond dimension is increased during imaginary time evolution as cutoff accuracy requires, which makes RPPS numerically more efficient than RMPS.

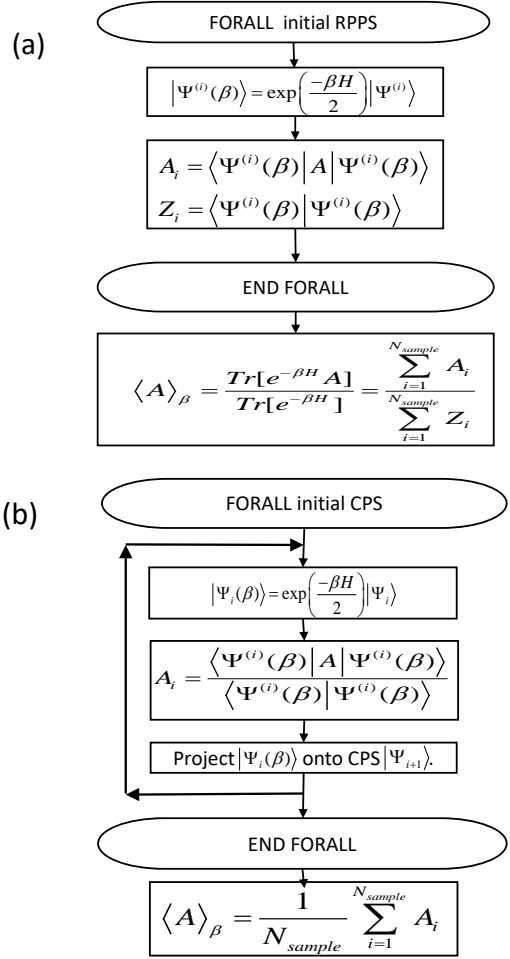

Fig. 3. Flow chart of (a) Random Phase Product State (RPPS) and (b) Minimally Entangled Typical Thermal State (METTS)[20, 21]. Sequential loop is designated with bold line.

Since the program for RPPS with Heisenberg chain model is written in C++ using hybrid-parallelization with OpenMP thread parallelization for ITensor library and MPI process parallelization for the initial RPPSs, the program will run extremely efficient on PC clusters and supercomputers such as FUGAKU. The source code will be submitted to the ITensor website[22] after the publication of this paper.

For large and almost homogeneous systems, macroscopic quantities calculated with random state method sometimes converge extremely rapidly and calculation with only one random state provides a good approximation. This super-convergence has been discussed in terms of *typicality* [7-12]. Derivation of RPPS method in this article also emphasizes the *self-averaging* [14, 19] due to the random phase on the off-diagonal matrix elements in (13).

In summary, we have introduced a numerical method that is useful for calculating canonical ensemble average in tensor network representation.

This research was supported by MEXT as "Exploratory Challenge on Post-K computer" (Challenge of Basic Science – Exploring Extremes through Multi-Physics and Multi-Scale Simulations). This research used computational resources of the Hokusai system provided by RIKEN and the HPCI system provided by JCAHPC through the HPCI System Research Project hp190167 and hp190196.